\documentclass[12pt]{article}
\usepackage{amsfonts}
\usepackage{amsmath}
\usepackage{amssymb}
\usepackage{graphics}
\begin{document}
\thispagestyle{empty}
 \newtheorem{theorem}{Theorem}[section]
 \newtheorem{proposition}[theorem]{Proposition}
 \newtheorem{corollary}[theorem]{Corollary}
 \newtheorem{lemma}[theorem]{Lemma}
 \newtheorem{defn}[theorem]{Definition}
 \newtheorem{remark}[theorem]{Remark}
 \newtheorem{example}[theorem]{Example}
\newcommand{\bcube}[1]{\mbox {$ \{0,1\}^{#1}$}}
\newcommand{\vect}[2]{\mbox {$ ({#1}_1, {#1}_2,\dots, {#1}_{#2}) $}}
\newcommand{\const}[1]{\mbox {$ \tilde {#1}$}}
\title{Combinatorial and Algorithmic Properties of One Matrix Structure
 at Monotone Boolean Functions}
\author{ Valentin Bakoev\thanks{Faculty of Mathematics and Informatics,
 University of Veliko Tarnovo, 2 Theodosi Tarnovski St, 5000
 Veliko Tarnovo, Bulgaria; email: v.bakoev@ts.uni-vt.bg}}

\date{}
\maketitle
\begin{abstract}
One matrix structure in the area of monotone Boolean functions is
defined here. Some of its combinatorial, algebraic and algorithmic
properties are derived. On the base of these properties three
algorithms are built. First of them generates all monotone Boolean
functions of $n$ variables in lexicographic order. The second one
determines the first (resp. the last) lexicographically minimal true
(resp. maximal false) vector of an unknown monotone function $f$ of
$n$ variables. The algorithm uses at most $n$ membership queries and
its running time is $\Theta(n)$. It serves the third algorithm, which
identifies an unknown monotone Boolean function $f$ of $n$ variables
by using membership queries only. The experimental results show that
for $1\leq n\leq 6$, the algorithm determines $f$ by using at most
$m.n$ queries, where $m$ is the combined size of the sets of
minimal true and maximal false vectors of $f$.
\end{abstract}
\vspace{2mm}
\\
{\small
{\it Note (Feb. 15, 2019).} This manuscript was written in 2005 and has not been published till now. This is its original version where few misprints have been corrected and the Internet references have been updated.
}
\section{Introduction}
\label{Intro}
The problems in the area of monotone Boolean functions (MBFs) are
important not only for the Boolean algebra. Many of them are related
to (or they have a direct interpretation in) problems, arising in
various fields, such as graph (hypergraph) theory, threshold logic,
circuit theory, artificial intelligence, computation learning
theory, game theory etc. \cite{AHK, MFLK, MI2}. Some problems,
concerning MBFs, are still not solved in the general case, others
have still open complexities. Some well-known scientists consider
that the capabilities of the known tools and methods for investigation
of MBFs are still not efficient enough and they recommend new
approaches and tools to be searched and used \cite{BHIK2, MI2}.
This opinion additionally motivated us to do the following
investigations and to represent them here.

Three of the well known problems, concerning MBFs are:

(1) The Dedekind's problem -- for enumeration of the MBFs of $n$
variables (or, equivalently, for enumeration of all antichains of
subsets of an $n$-set). The problem is set by Dedekind in the end
of 19th century, it is the oldest problem in the area of MBFs
\cite{Int2};

(2) The identification problem -- for identification of an unknown
MBF of $n$ variables by using a given learning model;

(3) Determining at least one minimal true vector and/or at least
one maximal false vector of an unknown MBF. This problem is closely
related to the problem (2).

Here we represent our investigations of these three problems.
Firstly, in Section \ref{Basic_Notions}, we recall some necessary
notions and known results. In Section \ref{The_Matrix} we define
one matrix structure, which represents the precedences of the
vectors of the $n$-dimensional Boolean cube. Some combinatorial
and algorithmic properties of the structure are derived. On the
base of them we build three algorithms. The first of them is
represented in Section \ref{Gen}. It generates all MBFs of $n$
variables in lexicographic order and it works in polynomial total
time. The second one has two versions and it is described in
Section \ref{LM1_RM0}. It determines the first (resp. the last)
lexicographically minimal true (resp. maximal false) vector of an
unknown MBF of $n$ variables. The algorithm is of the type binary
search, it has $\Theta (n)$ running time and uses at most $n$
membership queries for each of these vectors. Section
\ref{Identify} represents the third algorithm which identifies an
unknown MBF of $n$ variables by using membership queries only. It
uses the second algorithm and obeys to the "Divide and conquer"
strategy. Some comments concerning the realizations, the
complexity and the experimental results of the algorithm are also
given.
\section{Basic notions and preliminary results}
\label{Basic_Notions} One of the famous notions in Discrete mathematics is the {\it $n$-dimensional Boolean cube} \bcube{n} -- the $n$-th Cartesian power of the set \bcube{}, consisting of all $n$-dimensional binary vectors, i.e., \bcube{n}$\ =\{\vect{a}{n}|\, a_i\in \{0,1\},\ i=1, 2, \dots,n\}$. Obviously, these vectors are exactly $2^n$. The {\it serial number} of the vector $\alpha= (a_1, a_2$, $\dots, a_n)\in \bcube{n}$ is the natural number $\#\alpha= a_1.2^{n-1}+a_2.2^{n-2}+\dots+ a_n.2^0$, i.e., the natural number
whose binary representation is $a_1 a_2 \dots a_n$. The vector $\alpha \in \bcube{n}$ {\it precedes lexicographically} the vector $\beta\in \bcube{n}$ if either there exists an integer $k,1\leq k \leq n$, such that $a_k<b_k$ and $a_i=b_i$ for $i<k$, or $\alpha=\beta$. The vectors of \bcube{n} are in a {\it
lexicographic order} in the sequence $\alpha_0, \alpha_1,\dots, \alpha_{2^n-1}$, if $\alpha_i$ precedes lexicographically $\alpha_j$, for $0\leq i<j\leq 2^n-1$. When the vectors of \bcube{n} are in a lexicographic order (as we consider henceforth), their serial numbers form the sequence
$0,1,\dots,2^n-1$. The following inductive and constructive definition of the $n$-dimensional Boolean cube determines a procedure for obtaining its vectors in lexicographic order. 
\begin{defn}
\label{D10}
{\rm
\ 1) We call {\it one-dimensional Boolean cube} the set \bcube{}
$=\{(0),(1)\}$. Its elements $(0)$ and $(1)$ are one-dimensional
binary vectors and they are in lexicographic order.

2) Let $\bcube{n-1}=\{ \alpha_0, \alpha_1, \dots ,\alpha_{2^{n-1}-1}\}$
be the {\it $(n-1)$-dimensional Boolean cube} and let its elements, the
$(n-1)$-dimensional binary vectors $\alpha_0,\alpha_1,\dots,
\alpha_{2^{n-1}-1}$, be in lexicographic order.

3) We build the {\it $n$-dimensional Boolean cube} \bcube{n} by
\bcube{n-1}, firstly by adding 0 in the beginning of all its vectors,
and next by adding 1, i.e.,
$$
\bcube{n}=\{(0 \alpha_0),(0\alpha_1), \dots ,(0\alpha_{2^{n-1}-1}),
(1 \alpha_0),(1\alpha_1), \dots ,(1\alpha_{2^{n-1}-1}) \},
$$
and so the vectors of \bcube{n} are in lexicographic order.
}
\end{defn}

The relation "$\preceq$" is defined over $\bcube{n} \times \bcube{n}$ as follows: $\alpha \preceq \beta$ (we read "$\alpha$ precedes $\beta$") if $a_i\leq b_i$ for $i=1,2, \dots, n$. It is {\it reflexive, antisymmetric} and {\it transitive} and so \bcube{n} is a {\it partially ordered set} (POSet) with respect to the relation "$\preceq$". When $\alpha\preceq\beta$ or $\beta\preceq\alpha$ we call $\alpha$ and $\beta$ {\it comparable}, otherwise we call them {\it incomparable}.

The mapping $f : \bcube{n} \to \bcube{}$ is called a {\it Boolean
function} (or {\it function}, in short) of $n$ variables. If $\alpha,
\beta \in \bcube{n}$ and $\alpha \preceq \beta$ always implies
$f(\alpha)\leq f(\beta)$, then the function $f$ is called {\it monotone}
(or {\it positive}). We denote by $M_n$ the set of all MBFs of $n$
variables. When we consider the binary constants $0$ and $1$ as
functions, we denote them by \const{0} and \const{1}, respectively.
They are the unique functions of $0$ variables in $M_0$.

Let $f\in M_n$ and $\alpha \in \bcube{n}$. If $f(\alpha)= 0$ (resp.
$f(\alpha)=1$) then $\alpha$ is called a {\it false vector} (resp.
{\it true vector}) of $f$. The set of all false vectors (resp. all
true vectors) of $f$ is denoted by $F(f)$ (resp. $T(f)$). The false
vector $\alpha$ is called {\it maximal} if there is no other vector
$\alpha '\in F(f)$ such that $\alpha \preceq \alpha '$ and $\alpha
\neq \alpha '$. The set of all maximal false vectors is denoted by
$max\, F(f)$. Symmetrically, the true vector $\beta$ is called {\it
minimal} if there is no other vector $\beta '\in T(f)$ such that
$\beta '\preceq \beta$ and $\beta '\neq \beta$, also $min\, T(f)$
denotes the set of all minimal true vectors of $f$. Obviously,
each monotone function $f$ can be determined by only one of the
sets $min\, T(f)$ or $max\, F(f)$.

The function $x^\sigma$ is defined as follows: $x^\sigma=x$ if
$\sigma=1$, or $x^\sigma=\bar x$ if $\sigma=0$. The conjunction
$K=x_{i_1}^{\sigma_1}\dots x_{i_k}^{\sigma_k}$ is called an
{\it implicant} of the function $f$, if $T(K)\subseteq T(f)$.
If $K'$ and $K''$ are implicants of $f$, such that $T(K')\subset
T(K'')$, we say that $K''$ {\it absorbs} $K'$. The implicant
$K$ of $f$ is called {\it prime}, if there is not other implicant
$K'$ of $f$, such that $T(K)\subset T(K')$. The disjunction
$D=x_{j_1}^{\tau_1} \vee \dots \vee x_{j_r}^{\tau_r}$ is called
an {\it implicate} (or {\it clause}) of the function $g$ if
$F(D)\subseteq F(g)$. If $D'$ and $D''$ are implicates of $f$,
such that $F(D')\subset F(D'')$, we say that $D''$ {\it absorbs}
$D'$. The implicate $D$ of $g$ is called {\it prime} if there is
not other clause $D'$ of $g$, such that $F(D)\subset F(D')$.

In \cite{BHIK1, BHIK2, VGLK, MI1, MI2} it is shown that each
monotone function $f$ has an unique irredundant (minimal)
disjunctive normal form (IDNF), consisting of all prime implicants
of $f$, and also an unique irredundant conjunctive normal form
(ICNF), consisting of all prime implicates of $f$. In both forms
all literals are uncomplemented, so the IDNF and the ICNF of an
arbitrary monotone function are superpositions over the set $\{xy,
x\vee y,\const{0}, \const{1}\}$. The existence of a bijection
between the set of prime implicants in IDNF of $f$ and the set
$min\, T(f)$ is also noted -- each prime implicant $K_i= x_{i_1}
x_{i_2} \dots x_{i_k}$ corresponds to the vector $\alpha \in min\,
T(f)$ having ones in coordinates $i_1, i_2, \dots, i_k$ and zeros
in all the rest coordinates. Hence $\alpha$ is a characteristic
vector of $K_i$. Analogously, in \cite{VGLK, MI1} is shown the
existence of a bijection between the set of all prime implicates
in ICNF of $f$ and the set $max\, F(f)$ -- each prime implicate
$D_j=x_{j_1}\vee x_{j_2}\vee \dots \vee x_{j_r}$ in the ICNF of
$f$ corresponds to the vector $\beta \in max\,F(f)$ having zeros
in coordinates $j_1, j_2,\dots, j_r$ and ones in all the rest
coordinates. So $\beta$ is an anti-characteristic vector of
$D_j$.
\begin{example}
\label{EX10}
  Let us consider the function $f(x,y,z)=(0,0,1,1,0,1,1,1)\in M_3$,
for which we have:

1) The IDNF of $f$ is $f(x,y,z)=y\vee xz$, and $y$, $xz$ are its prime
implicants. They corresponds bijectively to the vectors $(0,1,0)$,
$(1,0,1)$ and so $min\, T(f)=\{(0,1,0)$, $(1,0,1)\}$;

2) The ICNF of $f$ is $f(x,y,z)=(x\vee y)(y \vee z)$, and $(x\vee y)$,
$(y \vee z)$ are its prime implicates. They corresponds bijectively to the
vectors $(0,0,1)$, $(1,0,0)$ and hence $max\,F(f)=\{(0,0,1),(1,0,0)\}$.
\end{example}
\section{One matrix structure and its properties}
\label{The_Matrix}
We shall represent the relation "$\preceq$" over the vectors of
\bcube{n} by a matrix.
\begin{defn}
\label{D20}
{\rm
      We define a {\it matrix of the precedences} $P_n=||p_{ij}||$
of dimension $2^n\times 2^n$ as follows: for each pair of vectors
$\alpha, \beta \in \bcube{n}$, such that $\#\alpha=i$, $\#\beta=j$,
we put $p_{ij}=1$ if $\alpha\preceq \beta$, or $p_{ij}=0$ otherwise.
}
\end{defn}

The rows and the columns of $P_n$ are numbered from $0$ till $2^n-1$,
in accordance with the numbers of the vectors in \bcube{n}.
\begin{theorem}
\label{T10}
      For $n=1$ the matrix $P_1$ is
{\small
$\left(
\begin{array} {ll}
1\ 1\\0\ 1\\
\end{array} \right)$.
}
For any integer $n>1$  $P_n$ is a block matrix of the form
{\small
$$
P_n=\left(
\begin{array}
{ll}P_{n-1}\ P_{n-1}\\O_{n-1}\ P_{n-1}\\
\end{array}
\right),
$$
}
where $P_{n-1}$ denotes the same matrix of dimension
$2^{n-1}\times 2^{n-1}$, and $O_{n-1}$ is the zero
matrix of dimension $2^{n-1}\times 2^{n-1}$. $P_n$
represents the precedences of the vectors of \bcube{n}
in accordance with Definition \ref{D20}.
\\
{\bf Proof.} {\rm We shall prove the theorem by an induction on $n$.

1) Obviously, for $n=1$ the matrix $P_1$ is of the given form and it
represents the precedences of the vectors $(0)$ and $(1)$ in \bcube{}.

2) We suppose that the theorem is true for the matrix $P_{n-1}$,
which represents the precedences of the vectors in \bcube{n-1} in
accordance with Definition \ref{D20}.

3) Following Definition \ref{D10}, $\forall\ \alpha \in \bcube{n}$
$\Rightarrow \alpha= (0,\gamma)$ or $\alpha= (1,\gamma)$, where $\gamma
\in \bcube{n-1}$. For arbitrary vectors $\alpha, \beta \in$ \bcube{n},
in dependence of this whether they begin with 0 or with 1, we
consider the following four cases:

i) $\alpha=(0,\gamma), \beta=(0,\delta)$, where $\gamma, \delta \in
\bcube{n-1}$. Let $\#\alpha=i$, $\#\beta=j$. Then $i=\#\gamma$,
$j=\#\delta$, $0\leq i, j\leq 2^{n-1}-1$, and also $\alpha \preceq\beta$
iff $\gamma \preceq \delta$. So, for any such $i, j$ the elements $p_{ij}$
of the matrix $P_n$ have the same values as the elements with the same
indices in the matrix $P_{n-1}$. Therefore the matrix $P_{n-1}$ is
placed in the upper left block (quarter) of $P_n$.

ii) $\alpha=(0,\gamma), \beta=(1,\delta)$, where $\gamma, \delta
\in \bcube{n-1}$. Then $\#\alpha=\#\gamma=i$, $0\leq i\leq 2^{n-1}-1$,
and $\#\beta=2^{n-1}+\#\delta=j$, $\ 2^{n-1}\leq j\leq 2^n-1$. Also
$\alpha \preceq\beta$ iff $\gamma \preceq \delta$. For these values of
$i$ and $j$ the elements $p_{ij}$ of the matrix $P_n$ are the same as
the elements $p_{ik},\ k=j-2^{n-1}$, of the matrix $P_{n-1}$.
So $P_{n-1}$ is placed in the right upper block of $P_n$.

iii) $\alpha=(1,\gamma), \beta=(0,\delta)$, $\gamma, \delta \in
\bcube{n-1}$. Every vector beginning with 1 does not precede a
vector beginning with 0. For the numbers of these vectors we have:
$i= \#\alpha$, $2^{n-1}\leq i \leq 2^n-1$ and $j= \#\beta$,
$0\leq j\leq 2^{n-1}-1$. Hence $p[i,j]=0$ for all such $i$ and $j$,
and so the zero matrix $O_{n-1}$ is placed in the left lower block
of $P_n$.

iv) $\alpha=(1,\gamma), \beta=(1,\delta)$,  $\gamma, \delta \in
\bcube{n-1}$. The case is analogous to the case (i), the difference
is only in the numbers of the vectors: $2^{n-1}\leq \#\alpha, \#\beta
\leq 2^n-1$. Analogously we prove that the matrix $P_{n-1}$ is
placed in the right lower block of $P_n$.

Therefore the matrix $P_n$ has the structure which states the theorem.
Also $P_n$ represents the precedences of the vectors of \bcube{n} in
accordance with Definition \ref{D20}, since the matrix $P_{n-1}$ do
this for the vectors of \bcube{n-1} (because of the inductive
suggestion). So the theorem is proved. $\diamond$
}
\end{theorem}
\begin{remark}
\label{Rem10}
{\it
From the properties of the relation "$\preceq$" and from the theorem it follows that $P_n$ is a triangular matrix, having ones on its major diagonal and zeros under it. The triangle of numbers on and over the major diagonal of $P_n$ is related to other known structures:

1) it is a discrete analog of the fractal structure known as
Sierpinski triangle;

2) the transposed matrix $P_n^T$ coincides with the Pascal's triangle consisting of $2^n$ rows, where the numbers are taken modulo 2, i.e., over $GF(2)$.
}
\end{remark}

The matrix $P_n$ can be expressed recursively as a Kronecker product:
$P_n=P_1\otimes P_{n-1}=P_1\otimes P_1\otimes P_{n-2}=\dots=$ Kronecker
$n$-th power of $P_1$.

We denote by $R_n=\{r_0,\dots, r_{2^n-1}\}$ the set of all rows
of $P_n$ considered as binary vectors.
\begin{theorem}
\label{T20}
Let $\alpha=(a_1,a_2,\dots, a_n) \in \{0,1\}^n,\ \#\alpha=i$,
$1\leq i\leq 2^n-1$, and $\alpha$ has ones in the coordinates
$i_1,i_2, \dots, i_r,\ 1\leq r\leq n$, i.e., $\alpha$ be the
characteristic vector of the conjunction $c_i=x_{i_1} x_{i_2}
\dots x_{i_r}$ (so it is a monotone function). If we consider
$c_i$ as a function of $n$ variables, then the vector of its
functional values contains the same values as (corresponds to)
the $i$-th row $r_i$ of the matrix $P_n$. When $\#\alpha=0$,
the zero row $r_0$ of $P_n$ corresponds to the \const{1}.
\\
{\bf Proof.} {\rm
We note, that we number the coordinates of the vectors of \bcube{n}
from left to the right, denoting by $x_1,x_2,\dots, x_n$ the variables
corresponding to them. Following Definition \ref{D10}, firstly we add
zeros and next we add ones in the beginning of each vector of \bcube{n-1}
to obtain the vectors of \bcube{n}. This is equivalent to an adding of
the variable $x_1$ in the beginning and increasing the indices of all
variables of \bcube{n-1} by one.

All elements of the zero row $r_0$ of $P_n$ are ones and so it
corresponds to the vector of \const{1}, as a function of $n$
variables. The rest part of the assertion we shall prove by
induction on $n$.

1) Obviously, the assertion is true for the matrix $P_1$.

2) We suppose that the theorem is true for the matrix $P_{n-1}$,
i.e., $\forall\ i$, $1\leq i\leq 2^{n-1}-1$, the vector of functional
values (or briefly "vector of the function" henceforth) of the
conjunction $c_i=x_{i_1} x_{i_2} \dots x_{i_r}$, which characteristic
vector is $\alpha \in \bcube{n-1},\ \#\alpha=i$, coincides with the
$i$-th row $r_i$ of the matrix $P_{n-1}$.

3) Let $\alpha \in \bcube{n-1}$, $\#\alpha=i=2^{i_1}+ 2^{i_2}+
\dots+ 2^{i_m}$, $1\leq i\leq 2^{n-1}-1$, $1\leq m \leq n-1$, and
so $\alpha$ has ones in the coordinates $i_1,i_2,\dots,i_m$. In
accordance with the inductive suggestion, the row $r_i$ of $P_{n-1}$
coincides with the vector of the function $c_i=x_{i_1} x_{i_2} \dots
x_{i_m}$. We consider two cases:

i) Let $\beta\in \bcube{n}$ be the vector, which is obtained by
adding 0 in the beginning of $\alpha$. Then $\#\beta=i$, it has
ones in the coordinates $i_1+1,i_2+1,\dots,i_m+1$ and so $\beta$
is a characteristic vector of the conjunction $c_i'=x_{{i_1}+1}
x_{{i_2}+1}\dots x_{{i_m}+1}$. Following Theorem \ref{T10}, the row
$r_i'$ of $P_n$ is obtained by writing the row $r_i$ of $P_{n-1}$
two times one after another (as a concatenation of strings). So
$r_i'$ coincides with the vector of a function of $n$ variables,
which is obtained by adding the fictitious variable $x_1$ to the
function $c_i$. Therefore the row $r_i'$ of $P_n$ contains
the functional values of $c_i'$.

ii) Let $\beta\in \bcube{n}$ be the vector, which is obtained by
adding 1 in the beginning of $\alpha$. Then $\#\beta=2^{n-1}+i=k$,
it has ones in the coordinates $1, i_1+1,\dots,i_m+1$ and so $\beta$
is a characteristic vector of the conjunction $c_k=x_1 x_{{i_1}+1}
\dots x_{{i_m}+1}$. Following Theorem \ref{T10}, the first half of
the row $r_k$ of $P_n$ is a row from the zero matrix $O_{n-1}$, and
its second half is the row $r_i$ of $P_{n-1}$. We consider
$r_k$ as a vector of function of $n$ variables. It is obtained by
adding (in conjunction) the essential variable $x_1$ to a function
of $n-1$ variables. On the first half of the vectors of \bcube{n} we
have $x_1=0$ and so the values in the first half of $r_k$ are zeros.
On the second half of the vectors of \bcube{n} we have $x_1=1$ and
so the values in the second half of $r_k$ are the same as these of
$r_i$. Therefore the row $r_k$ of $P_n$ contains the functional
values of the conjunction $c_k$.

So the theorem is proved. $\diamond$
}
\end{theorem}

We denote by $C_n$ the set of all conjunction of $n$ variables
without negations.
\begin{remark}
\label{Rem20}
{\it
The correspondence in Theorem \ref{T20} between the conjunction $c_i$
and its characteristic vector $\alpha\in \bcube{n}$, $\#\alpha=i$, is
a bijection $\varphi :\bcube{n}\to C_n$. Theorem \ref{T20} states the
relation between the set $C_n$ and the matrix $P_n$ -- this is the
bijection $\psi : C_n \to R_n$, the bijection between the formula
representation $c_i$ and the vector representation $r_i$
of each conjunction of $n$ variables without negations.
}
\end{remark}

Table \ref{tab:Ass_Th2} illustrates the assertion of Theorem \ref{T20} for $n=3$.
\begin{table}[ht]
	\centering
	\begin{tabular}{|c|c|c|l|}
	\hline
 $\alpha=(x_1,x_2,x_3)$ & $i=\#\alpha$ & $P_3$ & $c_i$ \\
	\hline
 	(0 0 0) & 0 & 1 1 1 1 \, 1 1 1 1 & $\tilde 1^{^{^{^{.}}}}$\\
 	(0 0 1) & 1 & 0 1 0 1 \, 0 1 0 1 & $x_3$\\
 	(0 1 0) & 2 & 0 0 1 1 \, 0 0 1 1 & $x_2$\\
 	(0 1 1) & 3 & 0 0 0 1 \, 0 0 0 1 & $x_2 x_3$\\
 	(1 0 0) & 4 & 0 0 0 0 \, 1 1 1 1 & $x_1^{^{^{}}}$\\
 	(1 0 1) & 5 & 0 0 0 0 \, 0 1 0 1 & $x_1 x_3$\\
 	(1 1 0) & 6 & 0 0 0 0 \, 0 0 1 1 & $x_1 x_2$\\
 	(1 1 1) & 7 & 0 0 0 0 \, 0 0 0 1 & $x_1 x_2 x_3$\\
	\hline
	\end{tabular}
	\caption{Illustration of the assertion of Theorem \ref{T20}, for $n=3$}
	\label{tab:Ass_Th2}
\end{table}

We consider the conjunction and the disjunction over binary vectors
as a bitwise operations. So the vector of an arbitrary $f\in M_n$
can be expressed as a linear combination $f(x_1,x_2,\dots,x_n)=
a_0 r_0\vee a_1 r_1\vee \dots\vee a_{2^n-1}r_{2^n-1}$, where the
coefficients $a_0, a_1,\dots, a_{2^n-1} \in \{0,1\}$, and the trivial
combination corresponds to $\tilde 0$. When $f(x_1,x_2,\dots,x_n)=
c_{i_1}\vee c_{i_2}\vee \dots\vee c_{i_k}$ is an IDNF of $f$, then
the corresponding to the prime implicants rows $r_{i_1}, r_{i_2}
\dots, r_{i_k}$ are pairwise incomparable (as binary vectors)
and the vector of $f$ is a result of $r_{i_1}\vee r_{i_2}
\vee \dots\vee r_{i_k}$.

Let us consider an arbitrary row $r_i$ of the matrix $P_n$ and let
the values in positions $i_1, i_2, \dots,i_k$ ($i=i_1< i_2< \dots<i_k$)
be ones. Then the set of vectors $\{\alpha_{i_1},\alpha_{i_2},\dots,
\alpha_{i_k}\} \subseteq \bcube{n}$ is actually the set $T(c_i)$. The
vector $\alpha_{i_1}= \alpha_i$ precedes all the rest vectors of
$T(c_i)$ and therefore $min\, T(c_i)=\{\alpha_i\}$.

Now let $j$ be the position of the rightmost zero in an arbitrary
row $r_i$ of $P_n$. Let us consider the $j$-th column of $P_n$ and let
the values in positions $j_1, j_2,\dots,j_m$ ($j_1< j_2<\dots<j_m=j$)
of this column be ones. This means that each vector from the set
$\{\alpha_{j_1}, \alpha_{j_2},\dots, \alpha_{j_m}\} \subseteq \bcube{n}$
precedes the vector $\alpha_{j_m}= \alpha_j$. Since the row $r_i$
corresponds to the vector of a monotone function, the zero in position
$j$ of $r_i$ implies zeros in positions $j_1, j_2, \dots,j_{m-1}$. Hence
$\{\alpha_{j_1}, \alpha_{j_2},\dots, \alpha_{j_m}\}\subseteq F(c_i)$ and
only the last of them $\alpha_j \in max\,F(c_i)$. So $|max\,F(c_0)|=0$,
and $|max\,F(c_i)|\geq 1$ if $i>0$. As we have noted, the vector
$\alpha_j$ corresponds bijectively to the clause $d_j$, which
anti-characteristic vector is $\alpha_j$.
\begin{example}
\label{EX20}
	Let us consider the row $r_3$ of $P_3$ (see Table \ref{tab:Ass_Th2}). It
corresponds to the conjunction $c_3=x_2 x_3$ and has ones in positions
3 and 7. Therefore $T(c_3)=\{(0,1,1), (1,1,1)\}$ and $min\, T(c_3)=
\{(0,1,1)\}$. The rightmost zero in $r_3$ is in position 6. The sixth
column of $P_3$ has ones in positions $\{0,2,4,6\}$. Therefore
$\{(0,0,0),$ $(0,1,0),(1,0,0),(1,1,0)\}\subseteq F(c_3)$ and $(1,1,0)
\in max\,F(c_3)$. Actually, $max\,F(c_3)= \{(1,0,1),$ $(1,1,0)\}$,
$\alpha_5=(1,0,1)$ corresponds to the clause $d_5=x_2$,
and $\alpha_6=(1,1,0)$ -- to the clause $d_6=x_3$.
\end{example}

The following assertion is symmetrical to Theorem \ref{T20}.
\begin{theorem}
\label{T30}
Let $\alpha=(a_1,a_2,\dots, a_n) \in \{0,1\}^n,\ \#\alpha=i$,
$0\leq i\leq 2^n-2$ and $\alpha$ has zeros in the coordinates
$i_1,i_2, \dots, i_r,\ 1\leq r\leq n$, i.e., $\alpha$ be the
anti-characteristic vector of the disjunction $d_i=x_{i_1}\vee
x_{i_2}\vee \dots \vee x_{i_r}$ (so it is a monotone function).
If we consider $d_i$ as a function of $n$ variables, then the
values in its vector are the negated values of the $i$-th column
of the matrix $P_n$. When $\#\alpha=2^n-1$, the negated values
of the last column of $P_n$ are the values of the vector of
\const{0}.
\end{theorem}

The proof is analogous to the proof of Theorem \ref{T20} and
we omit it.
\section{Generating MBFs of $n$ variables in lexicographic order}
\label{Gen}
In Section \ref{Intro} it was mentioned that the oldest problem in
the area of MBFs is the Dedekind's problem -- for enumerating MBFs
of $n$ variables (or, equivalently, for counting all antichains of
subsets of a given $n$-element set). The numerous efforts of the
researchers for solving this problem led up to obtaining a lot of
estimations for $|M_n|$ (from above and below) \cite{KOR81, Int3}. Now
an exact formula for the number of MBFs of $n$ variables, in the
general case, is not known. Till now this number was known for
$0\leq n \leq 8$ only. The number of MBFs, known for us,
is represented in the following table (see \cite{Int2} and the
sequence A000372 in \cite{SLO}).
\begin{table}[ht]
	\centering
	\begin{tabular}{|c|r|}
	\hline
 	$n$ & $|M_n|$ \\
	\hline
  	0  & 2 \\ 
	\hline
  	1  & 3 \\ 
	\hline
  	2  & 6 \\ 
	\hline
  	3  & 20 \\ 
	\hline
  	4  & 168 \\
	\hline
  	5  & 7\,581 \\ 
	\hline
  	6  & 7\,828\,354 \\
	\hline
  	7  & 2\,414\,682\,040\,998\\
	\hline
  	8  & 56\,130\,437\,228\,687\,557\,907\,788\\
	\hline
	\end{tabular}
	\caption{$|M_n|$, for $0\leq n\leq 8$}
	\label{tab:Nun_MBF}
\end{table}

An important problem in Combinatorics is the problem for
generating the elements of a given set in a definite order. In the
area of MBFs this problem has an additional meaning -- it is an
approach ({\it generating and counting}) for a partial solving the
Dedekind's problem. In \cite{Int1} a computer program (written in
C++) for generating MBFs of $n$ variables, $0\leq n\leq 7$, is
represented. The program is used to test sorting networks, and
also for solving the Dedekind's problem for these values of $n$.
It realizes an algorithm, described in \cite{Int3} and based on
the following property. Let $f, g\in M_{n-1}$, $f$ and $g$ be
given by their vectors, and let $f\preceq g$ (i.e., $f\vee g=g$).
If $h$ is the function, which vector is a concatenation of the
vectors of $f$ and $g$ (considered as strings), then $h\in M_n$.
So generating the functions of $M_n$ requires: (1) all functions
of $M_{n-1}$ to be generated and stored, and (2) to check whether
$f\preceq g$, for all pairs $f, g\in M_{n-1}$. These two
characterizations decrease the speed of the algorithm.

Here we propose an algorithm, called {\sc Gen}, for generating the
MBFs of $n$ variables. It was created in 1995, its initial purpose
was to compute the value of $|M_7|$. We have done a series of
optimizations and experiments and in 1999 we have generated about
$7\%$ of the function in $M_7$ for about 150 hours total time, on
several 200 MHz computers, performing independent subproblems (for
comparison, at this time the functions of $M_6$ was generated for
6 seconds on a 300 MHz computer). The principle "generating and
counting" turned out not so powerful for solving this problem, as
the mixed (analytical and computational) approach of Yovovic etc.
\cite{GKVY, VLYO}. When we got to know about their results, after
a comparison of our partial results with the asymptotic estimations
in \cite{KOR81, Int3}, and after evaluating the total time for the
generation, our attempts were canceled.

Algorithm {\sc Gen} {\it generates the vectors of the functions of $M_n$ in lexicographic order}, for given $n$. It is based on the matrix $P_n$ in the sense of Remark \ref{Rem20} and the explanations after it. Namely, if the row $r_i$ of $P_n$ has zero in position $j$, $i<j<2^n-1$, then the rows $r_i$ and $r_j$ are incomparable and therefore $f=r_i\vee r_j=c_i\vee c_j\,\in M_n$. After that, if the vector of $f$ has zero in position $k$, $i<k<2^n-1$, then it and the row $r_k$ are incomparable, so we can put $f=f\vee r_k=f\vee c_k\,\in M_n$, and so on. More precisely, the algorithm works with the rows of $P_n$ consecutively, starting from the last row. For $i=2^n-1, 2^n-2, \dots, 1, 0$, the algorithm puts $f=r_i$ and outputs it. Thereafter, while the vector of $f$ contains zeros in the positions after the $i$-th one, the algorithm does the following: it determines the position $j$ of the rightmost zero in $f$, performs the bitwise disjunction $f\vee r_j$, assigns the result to $f$ and outputs it. Thus the vectors of the functions of $M_n$ will be generated lexicographically. The formal description of the algorithm is:
\\
\\
{\bf Algorithm Gen.} Generates the MBFs of $n$ variables in a
lexicographic order.
\\
{\it Input:} $n$.
\\
{\it Output:} the vectors of the functions of $M_n$ in a
lexicographic order.
\\
{\it Procedure:}
\\
1) Put $f=\tilde 0$. Print $f$.\\
2) For each row $r_i$, $i=2^n-1, 2^n-2,\dots, 0$, put $f=r_i$ and:

a) print $f$;

b) for each $j$, $j=2^n-2, 2^n-3,\dots, i+1$, check the $j$-th
position of $f$. If $f[j]=0$, then put (recursively) $f=f\vee r_j$
and go to step a).
\\
3) End.
\\

The main part of the realization of {\sc Gen}, written in Pascal, is:
{\small
\begin{verbatim}
 1) Program Gen;
 2) .....
 3) Procedure Generate (G: BoolFun; i : integer);
 4)   var  j : integer;
 5)   begin
 6)     for j:= i to dim do      { Disjunction between the i-th row }
 7)       if P[i,j]=1 then  G[j]:= 1;  { and the current function G.}
 8)     Print (G);
 9)     for j:= dim-1 downto i+1 do        { Searching a zero for }
10)      if G[j]=0 then  Generate (G, j);  { the next disjunction.}
11)  end;   { Generate }

12) Begin   { Main }
13)   ...
14)   readln (n);            { Number of variables }
15)   dim:= 1 shl n - 1;     { dim:= 2^n-1 }
16)   Fill_Matrix;           { Filling in the matrix P_n }
17)   for k:= 0 to dim do  F[k]:= 0;  { Initialization - constant 0 }
18)   Print (F);
19)   for k:= dim downto 0 do  Generate (F, k);
20) End.
\end{verbatim}
}

{\bf Comments on the algorithm and its realization:}

1) The procedure {\sc Fill\_Matrix} in row 16) generates and stores
in the memory the matrix $P_n$ -- in accordance with either
Theorem \ref{T10}, or Remark \ref{Rem10}. In our realization (in
Borland Pascal 7.0) $1\leq n\leq 7$, and this restriction depends
on the realization only, it does not concern the nature of the
algorithm. When $n>7$ the matrix becomes too large -- then more
powerful program environment, or bitwise representation of the
elements of $P_n$ have to be used.

2) {\sc Gen} actually generates the functions lexicographically,
since the cycles in rows 9 and 19 have a decreasing step.

3) {\sc Gen} can generate only these functions, which come
lexicographically after the given function $h\in M_n$. For this
purpose, in row 17 we have to initiate $F=h$ instead of $F=\const{0}$,
and also the cycle in row 19 has to start from $i$ -- the position of
the leftmost one in the vector of $h$. Alternatively, if we change only
the final value in the same cycle -- for example, to be $m\,(0<m<dim)$
instead of 0, then {\sc Gen} will generate only these functions of
$M_n$, which precede lexicographically the function, which vector
is $r_{m+1}$.

4) The maximal number of rows of $P_n$, which are pairwise
incomparable determine the depth of the recursion in {\sc Gen}.
The disjunctions of these rows give a function, having a maximal
number of true vectors, namely $\displaystyle{\max_{f\in M_n}
|min\,T(f)|=}$ $\binom{n}{\lfloor n/2\rfloor}$ \cite{BHIK1, BHIK2}
(this is the size of the longest antichain in the POSet
\bcube{n}).

5) {\sc Gen} has an exponential time-complexity -- the nature of the
problem is such that the size of the output is always exponential
towards the size of the input. More precise classification of such
algorithms is given in \cite{DJMY}. {\sc Gen} uses only the last
generated function and a part of a certain row of $P_n$ to generate
the next function. It does this in an {\it incremental polynomial time},
i.e., the new function is generated in time, which is polynomial in the
combined size of the input, the last generated function and one row
of $P_n$. So the algorithm runs in a {\it polynomial total time}
\cite{DJMY}, i.e., its running time is a polynomial in the
combined size of the input and the output.

6) {\sc Gen} can be modified easily to generate all antichains of
a given POSet. For this purpose it is enough to build a new matrix
$P'_n$ representing the corresponding relation for each pair of
elements of the POSet.
\section{Determination of a minimal true and a maximal false vector of
an unknown MBF}
\label{LM1_RM0}

The problem for the {\it determination of at least one minimal
true vector and/or at least one maximal false vector of an unknown
monotone function} is an important problem in the area of MBFs
\cite{GAIN, KAT1, KOMI}. It is closely related to another
important problem -- for {\it identification} of such a function.
In the publications in Russian these problems are considered under
the assumption that an arbitrary function $f\in M_n$ is studied
by using some operator $A_f$, such that $A_f(\alpha)= f(\alpha)$
(i.e., returns the value of $f$ on $\alpha$), for $\alpha\in
\bcube{n}$ \cite{GAIN, KAT1, NSOK}. In the papers in English
the same problems are investigated in the terminology of {\it
computational learning theory}, which goals are to define and
study useful models of learning phenomena from an algorithmic
point of view \cite{DAN, AHK}. In the investigation of these two
problems the learning algorithm asks an oracle two types of
queries for an unknown function $f\in M_n$:

-- {\it membership queries} -- whether a selected vector $\alpha$
is a true (or a false) vector for $f$. The oracle answers "Yes" or
"No";

-- {\it equivalence queries} -- whether the unknown function $f$ is
equivalent with the hypothesis-function $g$. The oracle replies either
"Yes", or returns a counterexample (an arbitrary vector $\alpha$, such
that $f(\alpha)\neq g(\alpha)$).

The problem for determining at least one minimal true vector
and/or at least one maximal false vector of an unknown monotone
function is solved algorithmically. Effective algorithms, which
determine the corresponding vector(s) by using a minimal number of
membership queries and having a minimal time-complexity, are
searched (created) in the investigations. The considered problem
and its generalization (for $k$-valued monotone functions) are
studied by Katherinochkina \cite{KAT1}. The estimations, derived
by her show that an algorithm for determining an arbitrary maximal
false vector of an unknown $f\in M_n$ needs at least
$\binom{n}{\lfloor n/2 \rfloor}$ references to the corresponding
operator $A_f$. For an unknown $f\in M_n$, Gainanov \cite{GAIN}
proposes an algorithm which determines a new vector $\alpha$, such
that either $\alpha \in min\, T(f)$, or $\alpha \in max\, F(f)$.
The algorithm works as follows. Let $\alpha\in \bcube{n}$, its
coordinates $j_1,j_2, \dots,j_k$ be ones and $A_f(\alpha)=1$. Let
$e_i\in \bcube{n}$ be the $i$-th unit vector (only $i$-th its
coordinate is one, and all the rest are zeros). Algorithm builds
the sequence $f(\alpha^i), \,i=1,2,\dots,k$, such that

$\alpha^0=\alpha$,

$\alpha^i=\alpha^0\oplus f(\alpha^1)e_{j_1}\oplus\dots\oplus
f(\alpha^{i-1})e_{j_{i-1}}\oplus e_{j_i},\ i=1,2,\dots,k.$\\

Let $p$ be the maximal index for which $f(\alpha^p)=1$. Then the vector
$\beta=\alpha^p$ is a new vector for $min\, T(f)$. The case $A_f(\alpha)=0$
is treated analogously -- the coordinates $l_1,l_2,\dots,l_r$ of the zeros
in $\alpha$ are considered and the algorithm builds the sequence
$f(\alpha^i),\, i=1,2,\dots,r$, such that:

$\alpha^0=\alpha$,

$\alpha^i=\alpha^0\oplus \bar f(\alpha^1)e_{l_1}\oplus\dots\oplus
\bar f(\alpha^{i-1})e_{l_{i-1}}\oplus e_{l_i},\ i=1,2,\dots,r.$\\

If $q$ is the maximal index for which $f(\alpha^q)=0$, then the
vector $\beta=\alpha^q$ is a new vector for $max\, F(f)$.
The Gainanov's algorithm refers to the operator $A_f$ $O(n)$ times,
its time-complexity is $O(n)$. It is among the most effective algorithms
of this type. That is why it is so popular and useful for other
algorithms -- for example, in identification of monotone functions
\cite{BHIK1, BHIK2, MI2}.
\begin{defn}
\label{D30}
{\rm
      Let $f\in M_n$. The vector $\alpha\in min\, T(f)$ is called {\it
lexicographically first minimal true} vector (LFMT vector, in
short), if $\alpha$ precedes lexicographically each other vector
$\beta \in min\, T(f)$. Symmetrically, the vector $\gamma \in
max\, F(f)$ is called {\it lexicographically last maximal false}
vector (LLMF vector), if each other vector $\delta \in max\, F(f)$
precedes lexicographically $\gamma$. }
\end{defn}

Here we propose an algorithm in two versions, called {\sc
Search\_First} (resp. {\sc Search\_Last}), for determining LFMT
(resp. LLMF) vector of an unknown $f\in M_n$. The algorithm is
based on the block structure of the matrix $P_n$ and its
properties, given in Theorem \ref{T10}, Theorem \ref{T20} and the
explanations after it. The vector of an arbitrary $f\in
M_n\backslash \{\const{0}\}$ is a componentwise disjunction of
some incomparable rows of $P_n$. {\sc Search\_First} determines
the minimal number of a row among these in the disjunction -- so it
is the number of the LFMT vector of $f$. Symmetrically, {\sc
Search\_Last} determines the zero component in $f$, having a
maximal number -- so it is the number of the LLMF vector of $f$. In
both versions we use the variables {\sl left} and {\sl right},
denoting the left and the right limit (correspondingly) of the
interval for search. Their initial values are: $left= 0$ and
$right= 2^n-1$. {\sc Search\_First} asks membership queries for
the vectors of \bcube{n}, having numbers of the type
$m=(left+right)\ div\ 2$, i.e., whether $f(\alpha_m)=1$? If "Yes",
it puts $right= m$, otherwise it puts $left= m+1$. {\sc
Search\_First} computes the next value of $m$ (by the same
equality), it asks membership query for $\alpha_m$ again and
changes the value of either $left$, or $right$, an so on, until
the condition $left < right$ is true. In other words, the
algorithm performs simply a binary search. Theorem \ref{T10}
implies the correctness of this approach -- all rows from the
upper half of $P_n$ contain 1 in position $m=\lfloor
(0+2^n-1)/2\rfloor= 2^{n-1}-1$, and all the rows from the lower
half of $P_n$ contain 0 in the same position. The same is valid
for the blocks $P_{n-1}$ and the corresponding positions:
$m=2^{n-2}-1$ for the left upper block (i.e., after replay "Yes",
when $right$ is changed), or $m=2^{n-1}+ 2^{n-2}-1$ for the lower
right block (i.e., after replay "No", when $left$ is changed). And
so on, until some block $P_1$ is reached. Here is the code of {\sc
Search\_First}, written in Pascal.
{\small
\begin{verbatim}
Function Search_First (left, right : integer) : integer;
  var m : integer;
  begin
    while left < right do
      begin
        m:= (left + right) div 2;
        if f[m] = 1        { treated as a membership query }
          then right:= m
          else left:= m+1;
      end;
    Search_First:= left;
  end;
\end{verbatim}
}

{\sc Search\_Last} works in analogous manner, the membership
queries are used for the vectors, having numbers of the type
$m=((left+right) \ div\ 2)+1$, i.e., whether $f(\alpha_m)=0$? If
"Yes", the algorithm puts $left= m$, otherwise it puts $right=
m-1$, and so on. The explanations of the performance and the
correctness of {\sc Search\_Last} are analogous to these of the
previous one, its code is similar to the code of {\sc
Search\_First} and so we omit them.

\vspace*{2mm}
{\bf Comments on the algorithms and their realizations:}

1) {\sc Search\_First} determines the LFMT vector, and {\sc
Search\_Last} -- the LLMF vector of an unknown $f\in M_n \backslash
\{\const{0}, \const{1}\}$ by using $n$ membership queries, their
running time is $\Theta(n)$. When $f=\const{0}$ (resp.
$f=\const{1}$), the LFMT (resp. LLMF) vector does not exist and a
simple modification in the functions has to be done -- for example,
one more membership query has to be asked, or (equivalently): if
$f=\const{0}$ and {\sc Search\_Last} is started after {\sc
Search\_First}, then the first functions determines $\const{0}$
successfully by one additional query only (we use this approach in
identification).

2) The algorithm performs integer divisions only (for computing
indices), it does not generate any vectors and it is preliminarily
known what kind of vector will be searched and found --
contrary to the algorithm of Gainanov.

3) For clarity, in this section we represented a simplified
version of the algorithm, working on completely unknown function
$f$ (i.e., there is not partial knowledge about it). In its real
(extended) version, the algorithm asks membership query only when
the current vector is unknown (i.e., the value of $f$ on it cannot
be deduced by the current partial knowledge).
\begin{example}
\label{EX30}
	The following table illustrates the performance of the algorithm
on a sample function $f\in M_4$, treated as an unknown. The order of
the tested positions (of its vector) is shown in the third row of the
table. Those, tested by {\sc Search\_First}, are denoted with Arabic
numbers, and those, tested by {\sc Search\_Last}, are denoted with
roman numbers.
\begin{table}[ht]
	\centering
	{\small
	\begin{tabular}{|c|c|c|c|c|c|c|c|c|c|c|c|c|c|c|c|c|}
	\hline
	\# of a component &0&1&2&3&4&5&6&7& 8 &9&10&11&  12&  13& 14  &15\\
	\hline
	$f(x_1,x_2,x_3,x_4)^{^{^{.}}}$&0&0&1&1&0&0&1&1& 0 &1& 1& 1&   0&   1&  1  &1\\
	\hline
	order of testing & &3&4&2& & & &1&i& &  &  &ii&iv&iii&\\
	\hline
	\end{tabular}
	\caption{Illustration of the performance of {\sc Search\_First} and {\sc Search\_Last}}
	\label{tab:Ill_SF_and_SL}
}
\end{table}
\end{example}

We use this example to note, that if the real version of {\sc Search\_Last} (its code is given in the next section) is started after this one of {\sc Search\_First}, then its third membership query becomes unnecessary -- the value of $f$ in position 14 is defined by the prime implicant $c_2$, which is already determined by {\sc Search\_First}. These versions are included and controlled by an algorithm for identification, represented in the next
section. When the size of the vector decreases and/or there is some partial knowledge about the function, the necessary number of queries and the time-complexity of the algorithm decrease.

4) The performance of {\sc Search\_First} and {\sc Search\_Last} (in
their extended versions they register each new prime implicant/implicate,
which they find) has the following additional properties.
\begin{proposition}
\label{Prop10}
Each new prime implicant (resp. implicate), which the function
{\sc Search\_First} (resp. {\sc Search\_Last}) finds, absorbs the
previous one, found by the corresponding function. In the general case,
the same is not true for the set of implicates (resp. implicants),
which {\sc Search\_First} (resp. {\sc Search\_Last}) finds.
\end{proposition}
\begin{proposition}
\label{Prop20}
Let {\sc Search\_First} and {\sc Search\_Last} be executed one by
another on the function $f\in M_n$ and let the numbers of the determined
LFMT and LLMF vectors be $i$ and $j$, correspondingly. When the vector
of $f$:

a) is of size 4 (i.e., $f\in M_2$), or

b) is of the type $(0,\dots,0,1,\dots,1)$ (i.e., $i=j+1$), or

c) is of the type $(0,\dots,0,1,0,1,\dots,1)$ (i.e., $i+1=j$),\\
then the sets $min\, T(f)$ and $max\, F(f)$ are determined completely
in the process of searching.
\end{proposition}

The truth of these assertions follows directly from the given
explanations and comments about the performance of the algorithm
(see the order of the tested positions). They can also be proved
strongly by induction on $n$.
\section{Identification of an unknown MBF}
\label{Identify}
The problem for {\it Identification of an unknown MBF} $f\in M_n$
is also solved algorithmically, which means that the learning
algorithm determines the sets $min\,T(f)$ and $max\, F(f)$ as an
output -- so $f$ is specified completely. When the algorithm uses
membership queries only, this is an example of {\it exact learning
(of a Boolean theory $f$) by membership queries}, in the
terminology of computational learning theory \cite {AHK, BHIK2,
MI2}. When the learning model allows both membership and
equivalence queries to be asked, there are algorithms (for example
in \cite{DAN, GAIN}), which determine $min\, T(f)$ of an unknown
$f\in M_n$ by using $O(n|min\,T(f)|)$ membership and equivalence
queries generally. Here we consider and discuss only the first
learning model.

In \cite{BHIK2} are discussed some details, characteristics,
estimations and criteria for the learning algorithm, which uses
only membership queries. For such algorithm it is argued that:

1) it must determine both the sets $min\,T(f)$ and $max\, F(f)$,
although one of them can be obtained by another -- this requires
an exponential time (in the combined size of the both sets) it the
general case. Later, in \cite{VGLK} it is shown that the
determining of $max\, F(f)$ from $min\, T(f)$ is equivalent to
determining the set $min\, T(f^d)$, where $f^d\vect{x}{n}=\bar
f(\bar x_1, \bar x_2, \dots, \bar x_n)$ is the dual function
of $f$ (this problem is known as "{\it Dualization}" or "{\it
Transversal hypergraph}"). In \cite{VGLK} it is proved that this
problem can be solved in incremental quasi-polynomial time;

2) its complexity (i.e., the number of the asked queries and the time-complexity) has to be evaluated in the combined size of the input $n$ and the output $m= |min\,T(f)| + |max\,F(f)|$, including the time for generating the vectors for the queries. The size of $m$ can become as large as $\binom{n}{\lfloor n/2\rfloor}+ \binom{n}{\lceil n/2\rceil}$ and polynomiality in $n$ only can not be expected \cite{BHIK1, BHIK2}.

The existence of a polynomial total time algorithm for solving the
problem Identification is equivalent to the existence of such type
algorithms for solving many other interesting problems in areas as
hypergraph theory, theory of coteries, artificial intelligence,
Boolean theory \cite{JBTI, MI1}. These problems have still open
complexities in spite of the numerous investigations. The results
of Fredman, Khachiyan and Gurvich \cite{MFLK, VGLK} show that it
is unlikely these problems to be NP-hard.

Makino, Ibaraki, Boros, Hammer etc. have studied intensively the
complexity of the Identification problem \cite{BHIK1, BHIK2, MI1,
MI2}. It is closely related to the complexity of the problem for
{\it determining a new vector} for some of the sets $MT\subseteq
min\,T(f)$ and $MF\subseteq max\, F(f)$, representing the partial
knowledge about the unknown function in a current stage. The
authors solve a restricted problem, they propose some algorithms,
which decide whether the unknown function $f$ is 2-monotonic or
not, and if $f$ is 2-monototonic they identify it in polynomial
total time and by using polynomial number of queries. In
\cite{MI1, MI2} is introduced the notion {\it maximum latency} as
a measure of the difficulty in finding a new vector. It is shown
that if the maximum latency of the unknown function $f\in M_n$ is
a constant, then an {\it unknown vector} (i.e., vector for which
the partial knowledge is insufficient to decide a true or a false
vector is it) can be found in polynomial time and there is an
incrementally polynomial-time algorithm for identification (the
algorithm in \cite{MI2} uses $O(n^2m)$ time and $O(n^2m)$ queries).
In \cite{MI1} it is proved that restricted classes of monotone
functions have a constant maximum latency. On the base of these
results in \cite{SHKA} it is proved that almost all MBFs are
polynomially learnable by membership queries.

Here we propose an algorithm, called {\sc Identify}, which identifies an unknown $f\in M_n$ by membership queries only. It is based on the properties of the matrix $P_n$ and uses the algorithm for determining LFMT and LLMF vectors. We consider {\it the problem for identification as an opposite to the problem for generating the MBFs}, i.e., during the generation, the algorithm {\sc Gen} combines (by disjunctions) some incomparable rows of $P_n$, whereas in identification of such a function (considered as unknown) the learning algorithm has to determine (or to separate) these rows. The difficulty in this process is due to the fact that the rows have positions, where their elements coincide.

{\sc Identify} determines both $min\, T(f)$ and $max\, F(f)$ of an unknown $f\in M_n$, i.e., the knowledge about $f$ is $n$ (which is the input) and its monotonicity. The algorithm works recursively and it obeys to the following main idea. Firstly it determines the LFMT and the LLMF vector of $f$ by using {\sc Search\_First} and {\sc Search\_Last} (their extended versions). After that it splits $f$ into two subfunctions $g, h\in M_{n-1}$, so $g\preceq h$, and the vector of $f$ is a concatenation $f=gh$ of the vectors of $g$ and $h$ (considered as strings) -- as it was mentioned in Section \ref{Gen}. So the LFMT (resp. LLMF) vector of $f$, which is found, is a LFMT (resp. LLMF) vector of $g$ (resp. $h$). The algorithm continues by identification of $g$ and $h$ in
the same way as $f$, i.e., it determines the LLMF vector of $g$ and recursively identifies it by splitting it into two subfunctions (having the mentioned above properties), and then it determines the LFMT vector of $h$ and identifies it in the same way. The recursive splitting into subfunctions continues in accordance with Proposition \ref{Prop20}, i.e., until a subfunction of some of the types in it is obtained -- so it is identified. In addition to Proposition \ref{Prop20} we note, that if a subfunction of the type $f'= \const{0}h'$, or $f'=g'\const{1}$ is obtained, then the subfunction \const{0} (resp. \const{1}) is already identified and the algorithm continues with an identification of $h'$ (resp. $g'$) only. So {\sc Identify} is a representative of the "{\it Divide and conquer}" strategy. Its main idea can be seen in a most clear form in the following procedure {\sc Id}, written in Pascal.
{\small
\begin{verbatim}
Procedure Id (left, right, lm1, rm0  : integer);
{ left is the initial position, and right is the final   }
{ position of the vector of f (or some its subfunction). }
{ lm1 is the position of the LFMT,  rm0 is the position  }
{ of the LLMF vector of f (or some its subfunction).     }

   var m  : integer;   { Position of the split.           }
       p0,             { Position of the LLMF vector of g.}
       p1 : integer;   { Position of the LFMT vector of h.}
   begin
         { Tests the cases of Proposition 5.4: }
 1)  if right-left <= 3 then exit;  { case a), }
 2)  if lm1 > rm0 then exit;        { case b), }
 3)  if lm1 = rm0+1 then exit;      { case c). }

 4)  t:= rm0-1;     { Test for a subfunction of the form (000...01)^k.}
 5)  if (rm0+1 = right) and (GetFunValue (t)= -1) then
       begin
         inc (q);   { Counting the queries by q is included.}
         if F[t] = 0 then     { When unknown - membership query,}
           Reg_Clause (t)     { registers a new clause, or      }
         else Reg_Impl (t);   { registers a new implicant.      }
       end;

 6)  m:= (left+right) div 2;  { Computing the position of the split.}
 7)  if lm1 > m then
       begin                                { When f'=0h' - }
         Id (m+1, right, lm1, rm0);  exit;  { identifies h'.}
       end;

 8)  if rm0 <= m then
       begin                                { When f'=g'1 - }
         Id (left, m, lm1, rm0);  exit;     { identifies g'.}
       end;
                                     { In the rest cases, f'= g'h':     }
 9)  p0:= Search_Last (left, m);     { - searching the LLMF vector of g'}
10)  Id (left, m, lm1, p0);          {   and identification of g';      }
11)  p1:= Search_First (m+1, right); { - searching the LFMT vector of h'}
12)  Id (m+1, right, p1, rm0);       {   and identification of h'.      }
   end;   {Id}
\end{verbatim}
}

As the comments in the source show, the "if"-operators in rows 1), 2) and 3) test the conditions of Proposition \ref{Prop20} for finishing the identification. These in rows 7) and 8) test whether \const{0} or \const{1} is a subfunction of the current function -- then the identification continues with the rest half of it. The test in row 5) was not discussed till now. There are functions (subfunctions), which vector is of the form $(0,\dots,0,1)^k$, i.e., the vector $(0,\dots,0,1)$ is repeated $k$ times. In the worst case they can have only one prime implicant and two prime implicates, independently on $n$. If $f$ is such a function, then $m=|min\, T(f)|+|max\,F(f)|=3$ and the number of queries, necessary for its identification, can grow non-polynomially in $n$ and $m$. The test in row 5) checks the third position from right to the left, so it recognizes such functions by one additional membership query and prevents the number of queries to grow unnecessary. For example, without this test the algorithm identifies the function $f=(0,0,0,1)^8\in M_5$ by using $16$ queries, and after including the test the number of queries becomes $10$. The function {\sc GetFunValue} and the procedures {\sc Reg\_Impl} and {\sc Reg\_Clause} in {\sc Id} are discussed in the following comments.

\vspace*{2mm}
{\bf Comments on the algorithm and its realizations:}

1) We wrote several versions of {\sc Identify} and we done many
experiments as well, trying to minimize the number of queries and
the time-complexity. Since our first goal was to minimize the number
of queries, the older versions generate the matrix $P_n$ and use a
partial function (i.e., a hypothesis-function $h$ of $n$ variables,
the values of its vector are marked as unknown initially; the
algorithm asks queries about $f$ and registers the obtained
knowledge in the vector of $h$, until it contains unknown values --
thereafter the $h$ is completely specified and $h=f$), similarly to
the algorithms in \cite{BHIK1, BHIK2, MI1, MI2}. This approach
always implies an exponential time-complexity. Generating and using
the matrix $P_n$ is justified when all functions of some large
enough set has to be identified.

2) In the last version we do not generate the matrix $P_n$. Instead
of this, the following function {\sc GetP} determines and returns the
value in the cell $p[i,j]$ of the matrix $P_n$ (the values of the
array $d$ are set $d[k]=2^k-1$, for $k=0,1,\dots, n$, initially).
{\small
\begin{verbatim}
Function GetP (i, j, m : integer) : boolean;
{ Determines the value in the cell p[i,j] of the matrix P_m.   }
{ GetP returns "true" when p[i,j]=1, or "false" when p[i,j]=0. }
  begin
    GetP:= false;
    while m>=1 do
      begin
        if i>j then   { If p[i,j] is under the major diagonal }
          exit;       { of P_m - returns "false".             }
        if (i=j) or (m=1) then  { If p[i,j] is on the major         }
          begin                 { diagonal of P_m, or p[i,j] is     }
            GetP:= true; exit;  { over this of P_1 - returns "true".}
          end;
        m:= m-1;  { Checks in which block of P_m is the cell p[i,j]:}
        if i > d[m] then    { if it is in the lower half,        }
          i:= i-d[m]-1;
        if j > d[m] then    { if it is in the right half,        }
          j:= j-d[m]-1;
      end;                  { the corresponding block is chosen. }
  end;   { GetP }
\end{verbatim}
}

Theorem \ref{T10} implies the correctness of the function {\sc
GetP}. The cycle "while" in it is repeated at most $m$ times, hence
the time-complexity of {\sc GetP} is $O(n)$ for $P_n$ (of dimension
$2^n \times 2^n$). The recursive version of {\sc GetP} is more
compact but it runs a bit slower.

3) The second main change in the last version of {\sc Identify} is
the discarding of the hypothesis-function. It may contain many
unnecessary values (for example, these before (resp. after) the
first LFMT (resp. last LLMF) vector), many checks and fillings
(sometimes of the type $2^n$) have to be done for registration of
each new prime implicant/implicate, which leads to an exponential
time-complexity. When {\sc Identify} needs to know the value of
$f$ in its $k$-th position (i.e., on $\alpha_k \in \bcube{n}$), it
tries to derive it from the partial knowledge about $f$. Let us
denote by $TPI (f)$ (resp. $TPC (f)$) the set of all temporary
prime (between each other) implicants (resp. clauses) of some
subfunctions of $f$, which are known at a current stage. In
accordance with Theorem \ref{T20} and Theorem \ref{T30}, the
algorithm performs:

a) checks the set $TPI (f)$: if for some $c_i\in TPI (f)$, $p[i,k]= 1$,
i.e., $GetP (i, k, n)$ returns "true", then the value in $k$-th position
is 1, otherwise it goes to step b);

b) checks the set $TPC (f)$: if for some $d_j\in TPC (f)$, $p[k,j]= 0$,
i.e., $GetP (k, j, n)$ returns "false", then the value in $k$-th position
is 0, otherwise it is unknown.

These two steps are realized by the function {\sc GetFunValue
}($k$), which returns the value of $f$ in its $k$-th position: $0$,
$1$, or $-1$ if it is unknown. So the time-complexity of {\sc
GetFunValue} will be $O(n(|TPI (f)|+|TPC (f)|))$. Although the set
$TPI (f)$ ($TPC (f)$) contains only prime (between each other)
implicants (clauses), we do not succeeded to estimate how large
can they become. Our experimental results show, that there are
only few functions in $M_6$, for which the maximal size of $TPI$
(or $TPC$), reached in identification, exceeds $m$ of the
corresponding function by one. Obviously, when the identification
finishes, then $TPI (f)=min\, T(f)$ and $TPC (f)=max\, F(f)$.

4) {\sc Identify} analyzes the answers to each membership query
and accumulates this partial knowledge by procedures, called {\sc
Reg\_Implicant} and {\sc Reg\_Clause}. Their parameter, the number
$k$, is of the just found new implicant or clause. In accordance
with Theorem \ref{T20} and Theorem \ref{T30}, the data
representation of the implicants and the implicates consists of
their numbers only, which are stored in two arrays, corresponding
to the sets $TPI (f)$ and $TPC (f)$. Following Proposition
\ref{Prop10}, each new implicant/clause is not compulsory prime,
so its registration consists of: (1) including it to the
corresponding array, and (2) excluding from the array of these
elements, which precede (are absorbed by) the new one. Step (1)
runs in a constant time, step (2) uses the function {\sc GetP} to
check the precedences at most $|TPI (f)|$ (resp. $|TPC (f)|$)
times. So the time complexity of {\sc Reg\_Implicant} (resp. {\sc
Reg\_Clause}) is $O(n|TPI (f)|)$ (resp. $O(n|TPC (f)|)$.

5) {\sc Identify} uses the real versions of {\sc Search\_First}
and {\sc Search\_Last}. The algorithm executes them to determine
the LFMT and the LLMF vector of $f$ before calling the procedure
{\sc Id}. Here is the real code of {\sc Search\_Last}.
{\small
\begin{verbatim}
Function Search_Last (left, right : integer) : integer;
  var m : integer;
      found : boolean;
  begin
    found:= false;
    while left < right do
      begin
        m:= (left+right) div 2 +1;
        case GetFunValue (m) of  { Checks the value of f in m-th position.}
          0  : left:= m;
          1  : right:= m-1;
          -1 : begin             { When this value is unknown -  }
                 if f[m]=0 then  { membership query:             }
                   begin         { - a new prime clause is found;}
                     left:= m;  found:= true;
                   end
                 else
                   begin         { - a new implicant is found.   }
                     right:= m-1;  Reg_Impl (m);
                   end;
                 inc (q);  { Counting the queries by q is included.}
               end;
        end;
      end;
    Search_Last:= right;
    if found then Reg_Clause (right);
  end;   {Search_Last}
\end{verbatim}
}

As the previous version of {\sc Search\_Last}, this one asks at
most $n$ membership queries to determine the LLMF vector of the
function $f$, which vector is of size $2^n$ (in fact, the real
number of the queries can be quite smaller because of the partial
knowledge). The time-complexity of this version is determined by
the cycle "while", which is executed exactly $n$ times. Then the
function {\sc GetFunValue} is executed $n$ times, the function
{\sc Reg\_Impl} is executed at most $n-1$ times, and the function
{\sc Reg\_Clause} is executed once. So the time complexity of
{\sc Search\_Last} is: $O(n.O(n(|TPI (f)|+|TPC (f)|))+(n-1).O(n|TPI
(f)|)+O(n|TPC (f)|)=$ $O(n^2(|TPI (f)|+|TPC (f)|))$. The real code
of {\sc Search\_First} is analogous to this one of {\sc Search\_Last}
and it has the same complexity as {\sc Search\_Last}. If we register
only the prime implcants/implicates, which they find, then always
$TPI (f)\subseteq min\, T(f)$ and $TPC (f)\subseteq \max\, F(f)$
and their sizes will be bounded by $m$. But the algorithm will ask
membership queries for one and the same vectors more than once and
the number of queries will grow extremely.

These versions of {\sc Search\_First} and {\sc Search\_Last} do not
generate any test-vectors. They compute the numbers of these vectors
only in a constant time.

\vspace*{2mm}
{\bf Comments on the complexity of the algorithm}

For the recent version of {\sc Identify} we do not succeeded to
estimate the number of queries, used in the general case. We have
considered some reasons (concerning the performance of {\sc
Search\_First} and {\sc Search\_Last} and its features), to assume
that the number of queries is polynomial in $n$ and $m$, but we
cannot prove (or disprove) this. We note that each LFMT and LLMF
vectors of the function/subfunction, which the algorithm finds,
bring maximal information (towards any other vectors) about it.
Also both functions ask membership query only in a case of
necessity, i.e., when the value of the unknown function on a given
vector cannot be derived from the current partial knowledge about
it and its monotonicity.

In the recent version of the algorithm we removed some reasons,
which lead to an exponential time-complexity. The assertions of
Proposition \ref{Prop20} and some other reasons, considered above,
simplify and speed-up the identification of many functions. So the
time-complexity of the algorithm decreases (towards the previous
versions), but it remains exponential in the general case. The
worst cases for the algorithm are such functions, where the
recursion stops only when a subfunction of two variables is
reached. Obviously, they contain subfunctions, which are one and
the same, but the algorithm does not recognize them -- so it
identifies each of them separately and independently. We note,
that in such cases some subfunctions can be determined completely
by the partial knowledge about the identified ones. Then {\sc
Search\_First} and {\sc Search\_Last} do not ask a new query. For
example, 17 functions in $M_3$ are identified in the process of
determining their LFMT and LLMF vectors, but the rest three
functions have an unknown vector -- so they need either {\sc
Search\_First} or {\sc Search\_Last} to be executed one more time.

\vspace*{2mm}
{\bf Experimental results}

We have done many experiments for identification of all functions
in $M_n, 0\leq n\leq 6$, treated as unknown. They are generated
preliminarily by the algorithm {\sc Gen} and are written in a file,
the prime implicants of each function are also written. For example,
on a 2.4 GHz processor and for $n=6$, the running time for generating
these functions is 9 minutes and 2 seconds. For the same parameters,
the identification (including some checks and collecting data for
statistics) continues 26 minutes and 17 seconds. The experimental
results show that for $0\leq n\leq 6$, the functions of $M_n$ are
identified by no more than $nm$ queries (the unique exception is
\const{0}, which is identified by $nm+1$ queries). Table \ref{tab:Num_of_Queries} represents the maximal $q_{max}$ and the average $q_{ave}$ number of membership query, used for identification, in dependence of $n$.
\begin{table}[ht]
	\centering
	\begin{tabular}{|c|c|c||c|c|c|}
	\hline
	$n$ & $q_{max}$ & $q_{ave}$ & $n$ & $q_{max} $ & $q_{ave} $ \\
	\hline
 	1  &  2  &  1.66  &  4  &  12  &  8.95\\
	\hline
 	2  &  3  &  2.66  &  5  &  22  & 16.76\\
	\hline
 	3  &  6  &  4.70  &  6  &  41  & 30.65\\
	\hline
	\end{tabular}
	\caption{Maximal and average number of queries, used for identification of the functions in $M_n$, for $1\leq n\leq 6$}
	\label{tab:Num_of_Queries}
\end{table}

\begin{figure}[!ht]
	\centering
  	\scalebox{1}{\includegraphics {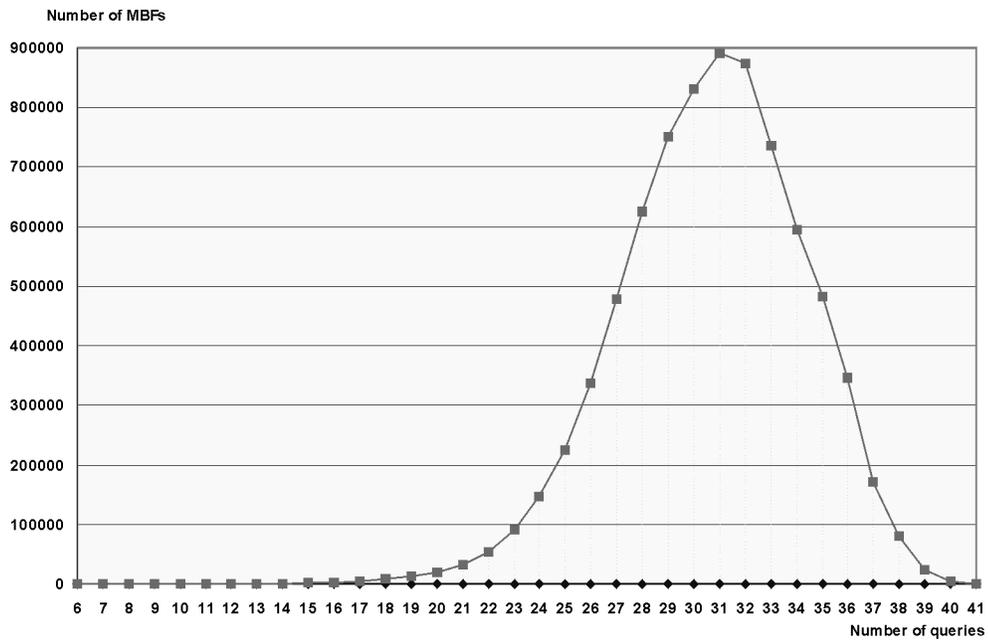}}
	\caption{Results of identification of all MBFs of 6 variables}
	\label{fig:Res_MBF6}
\end{figure}

\begin{figure}[!ht]
	\centering
  	\scalebox{1}{\includegraphics {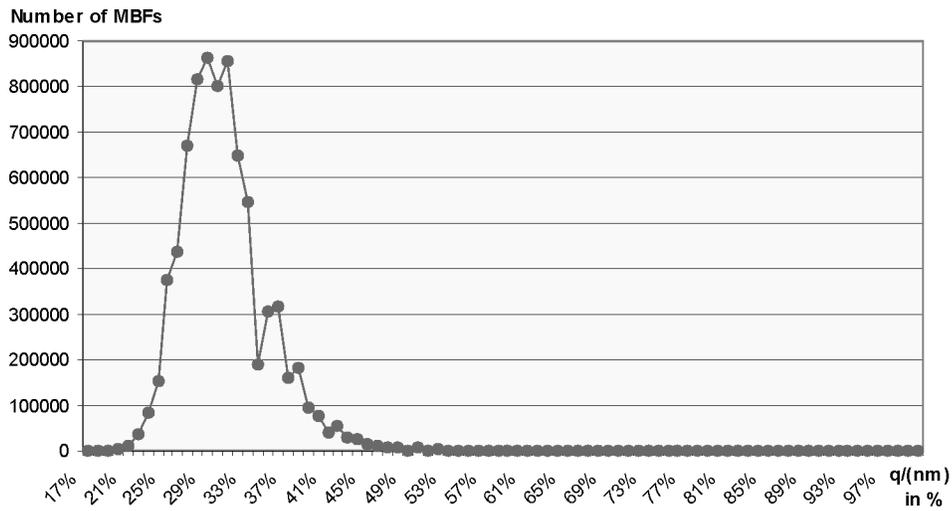}}
	\caption{MBFs of 6 variables with one and the same ratio $q/(nm)$ (in \%) obtained in their identification}
	\label{fig:Ratio_MBF6}
\end{figure}

Figure \ref{fig:Res_MBF6} and Figure \ref{fig:Ratio_MBF6} illustrate results of the identification of all MBFs in $M_6$. The diagram on the first figure represents how many MBFs are identified by the corresponding number of queries. The second diagram represents how many MBFs have one and the same ratio $q/(nm)$ (in \%), where $q$ is the number of asked queries, and
$m=|min\, T(f)|+ |max\, F(f)|$ is obtained in identification, for each $f\in M_6 \backslash \{\const{0}\}$.
\section{Conclusions}
\label{Concl}
In this work we represented our investigations on three important
problems, concerning MBFs. We introduced one matrix structure and
derived some of its combinatorial and algorithmic properties. They
were used as a base for building three algorithms, which was the
main reason these problems to be considered from a common point of
view. Solving the second and the third problems set some
questions, concerning the complexity of the corresponding
algorithms, which remained open. They will be subject to our
future investigations. We believe that the proposed approach,
matrix structure and algorithms have more (and better) properties
and capabilities than these, which we succeeded to obtain and
represent here.

\end{document}